\documentclass[fleqn,11pt]{olplainarticle}

\usepackage{timet}

\usepackage{amsmath}
\usepackage{amsfonts}
\usepackage{tabularx}
\usepackage{graphicx}
\graphicspath{{figs/}}
\usepackage{physics}
\usepackage[normalem]{ulem}
\usepackage{cancel}
\usepackage{xcolor}

\usepackage{tikz}
\usepackage{url}

\tikzset{>=latex}

\title{Joining simplified physics models with coarse grids to speed-up intractable 3D time-domain simulations}

\author[Wouter Deleersnyder]
{Wouter Deleersnyder$^{1,2}$, Evert Slob$^{3}$\\
	$^{1}$KU Leuven Campus Kortrijk - KULAK, Department of Physics, Etienne Sabbelaan 53, 8500 Kortrijk, Belgium.\\ 
	$^{2}$Ghent University, Department of Geology, Krijgslaan 281 - S8, 9000 Gent, Belgium \\  E-mail: {wouter.deleersnyder@ugent.be} \\
	$^{3}$ Delft University of Technology, Faculty of Civil Engineering and Geosciences , Department of Applied Geophysics and Petrophysics, Delft, Netherlands.
}

\begin{abstract}
Full 3D modelling of time-domain electromagnetic data requires tremendous computational resources. Consequently, simplified physics models prevail in geophysics, using a much faster but approximate (1D) forward model. We propose to join the accuracy of a 1D simplified physics model with the flexibility of coarse grids to reduce the modelling errors, thereby avoiding the full 3D accurate simulations. We exemplify our approach on airborne time-domain electromagnetic data, comparing the modelling error with the standard 3\% measurement noise. We find that the modelling error depends on the specific subsurface model (electrical conductivity values, angle representing the deviation of the 1D assumption) and the specific (temporal) discretization. In our example, the computation time is decreased by a factor of 27. Our approach can offer an alternative for surrogate models, statistical relations derived from large 3D datasets, to replace the full 3D simulations. 
\end{abstract}

\begin{document}

\flushbottom

\maketitle
\thispagestyle{empty}

\section{Introduction}
Proper characterization of the subsurface is crucial to groundwater management, CO$_2$ and H$_2$ storage, geothermal energy, and finding critical minerals for the energy transition. These challenges require increasingly advanced methods and models. For example, simpler analytical approximate models are being replaced by complex 3D simulations on high-performance computing infrastructure, yielding larger computational costs and a significantly larger energy footprint. In this study, we propose an alternative to those computationally expensive 3D simulations for  time-domain airborne electromagnetics (AEM), which require significant further processing of large datasets.\\

AEM is increasingly used for hydrogeological mapping \citep{mikucki2015deep, podgorski2013processing, ikard2023model}, saltwater intrusion \citep{siemon2019automatic, deleersnyder2023flexible} and contamination \citep{schamper2014assessment, pfaffhuber2017delineating}, and is still common for mineral exploration \citep{yang20173d, leycooper2022acquisition}. Those AEM systems have advanced massively in recent years \citep{auken2017review}, however, the data interpretation process and the associated computational burden remains a consequential impediment.\\

Full 3D inversion is an active research domain. 3D forward modelling is becoming applicable in the frequency domain for smaller areas \citep{ansari2017gauged, cockett2015simpeg,liu2017wavelet, werthmuller2019emg3d, rulff2021efficient, castilloreyes2023meshing}, but remains impractical for time-domain applications \citep{borner2015three, haber2007inversion, engebretsen2022accelerated}. It requires specialists’ expertise and a tremendous amount of resources. For example, \citet{engebretsen2022accelerated} could carry out a 2D time-domain electromagnetic (TDEM) inversion with 418 soundings, which took 30 hours, 97 GB of memory for only 16 iterations. Common AEM surveys are three dimensional and record $50,000$+ soundings. Consequently, pseudo/quasi-3D inversion methods are prevailing, using a much faster but approximate (1D) forward model. Alternative approaches were proposed in the past, based on the Born approximation, to generate multidimensional responses without numerical simulations \citep{oldenburg1991inversion, wolfgram2003approximate}, but this was not successful for inversion of AEM data \citep{guillemoteau2012fast}. Other methods used empirically inspired sensitivity functions \citep{christensen2002generic}, or approximate the sensitivity function with the Biot-Savart law. Each approach resulted a commendable speed-up, but they could not match the accuracy of numerical simulations in this decade. Recently, with the expansive adoption of artificial intelligence, so-called surrogate models, which are statistical models that approximate the simulation output and are faster to evaluate. The existing literature is already significant for 1D forward modelling \citep{bording2021machine, puzyrev2021inversion, asif2022integrating, wu2023deep}, reporting speed-ups of factor 50 to 2700, but requiring 100~000 to 1~000~000 training samples, posing significant challenges in 3D with a more significant parameter space and increased computational cost of data generation. In \citet{deleersnyder2024multidimensional}, a first attempt for 3D time-domain AEM was conducted, with only 5000 data samples and a limited two-layered case with a linear interface. The main limitation was the required numerosity of the training data to further generalize the approach.\\

In this work, we avoid expensive 3D simulations with a novel approach that joins the 1D analytical forward model with simulations on a coarse, computationally reasonable 3D grid. It combines the accuracy of the magnetic data in a simplified 1D case with the flexibility of numerical simulations. 

\section{Methods}
\subsection{Three types of forward models}
The forward model $\mathcal{F}$ for electromagnetic surveys describes the subsurface’s response $\vb{d}$ in the presence of a magnetic dipole source and known subsurface $\vb{m}$, which is a parametrized spatial electrical conductivity distribution.\\

There are two main forward modelling approaches: The first is based on (semi-)analytical models $\mathcal{F}_{\footnotesize  \mbox{1D}}$ that solve the (continuous) Maxwell equations for a one-dimensional model $\vb{m}_{\footnotesize \mbox{1D}}$, meaning that it assumes horizontal layers without lateral variations. An open-source Python implementation by \citet{werthmueller2017open} neatly implements such a forward model by \citet{hunziker2015electromagnetic} in a fast and reliable fashion. We refer to this model as the low-fidelity model (LF), as it cannot account for lateral variations in the subsurface model. The second approach discretizes the physics on a mesh, resulting in solving large matrix inverse problems. Those simulations mimic the full 3D soil response of the potentially non-1D subsurface and allow for multidimensional modelling, assuming a sufficiently fine discretization. In the case of perfect forward modelling $\mathcal{F}_{\footnotesize  \mbox{3D, HF}}$, we refer to these simulations as the high-fidelity model (HF).\\

\begin{figure}
	\noindent
	\begin{tabularx}{\textwidth}{rXp{0.01cm}Xll}
		
		& \bf \large \hspace{10mm} Input ($\vb{m}$) &~&\bf \large Forward Model ($\mathcal{F}$) & ~&\bf \large Output   \\
		\parbox{1em}{\vspace{-60pt}A.}&\noindent\parbox[b]{\hsize}{\includegraphics[width=100pt]{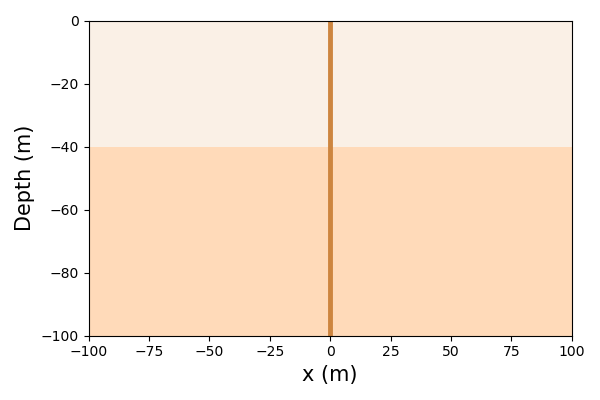} } &\parbox{0em}{\vspace{-60pt}\large$\to$ } &
		\includegraphics[width=100pt]{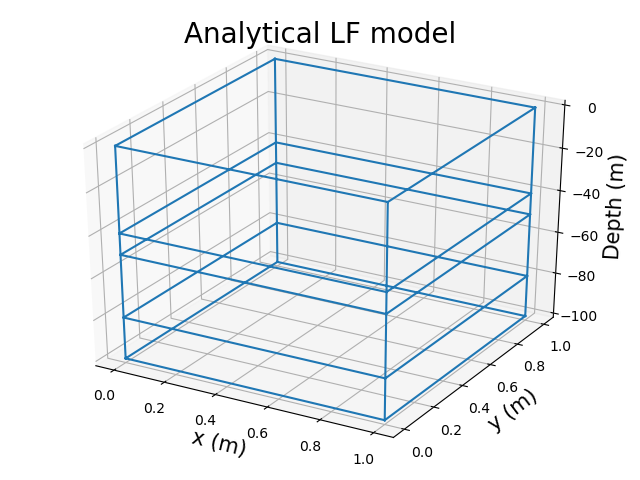} &\parbox{1em}{\vspace{-60pt}\large=}& \parbox{0em}{\vspace{-60pt}$\mathcal{F}_{\text{1D}}(\vb{m}_{\footnotesize \mbox{1D}})$} \\
		
		\parbox{1em}{\vspace{-60pt}B.}& \includegraphics[width=100pt]{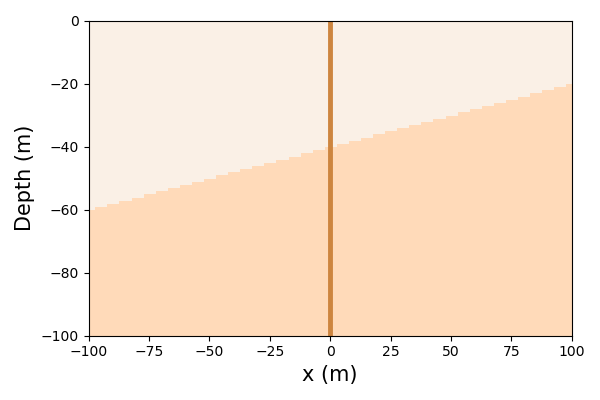}&\parbox{0em}{\vspace{-60pt}\large$\to$ }&\includegraphics[width=100pt]{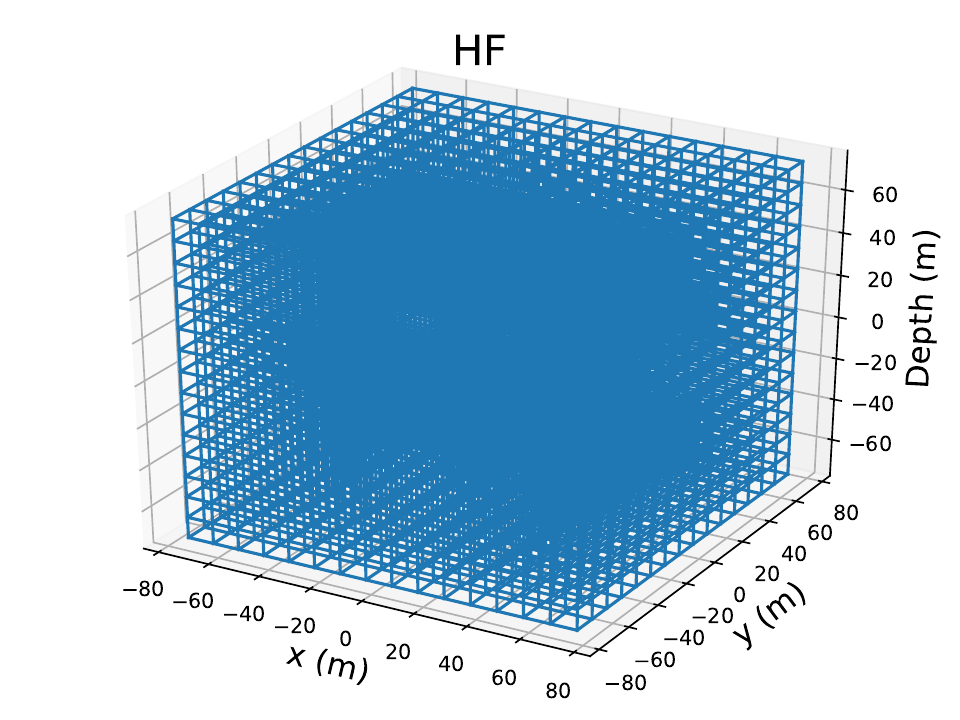} &\parbox{0em}{\vspace{-60pt}\large=}& \parbox{0em}{\vspace{-60pt}$\mathcal{F}_{\mbox{\footnotesize 3D, HF}}  (\vb{m})$}\\
		
		\parbox{1em}{\vspace{-60pt}C.}&\includegraphics[width=100pt]{figs/model_sub_1.png}&\parbox{0em}{\vspace{-60pt}\large$\to$ }  &\includegraphics[width=100pt]{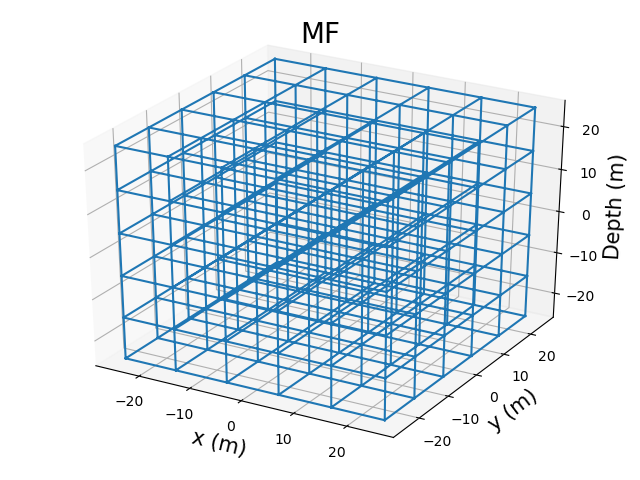} &\parbox{1em}{\vspace{-60pt}\large=}& \parbox{0em}{\vspace{-60pt}$\mathcal{F}_{\mbox{\footnotesize 3D, MF}}  (\vb{m})$} \\
		
		\parbox{1em}{\vspace{-60pt}D.} &\includegraphics[width=100pt]{figs/model_constant_1.png}&\parbox{0em}{\vspace{-60pt}\large$\to$ } &\includegraphics[width=100pt]{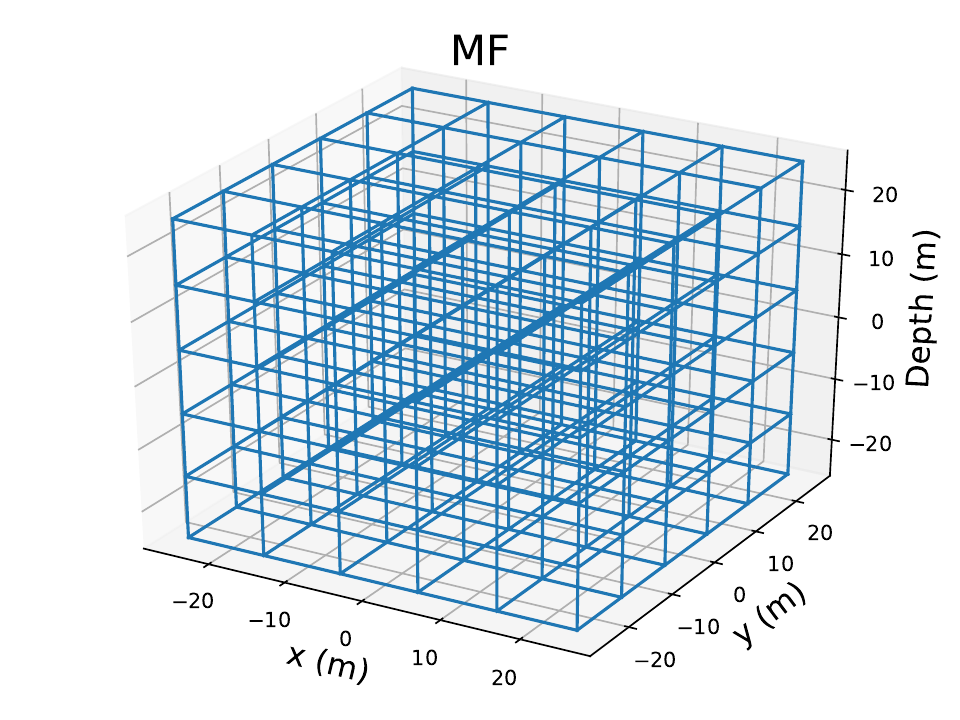} &\parbox{0em}{\vspace{-60pt}\large=}& \parbox{0em}{\vspace{-60pt}$\mathcal{F}_{\mbox{\footnotesize 3D, MF}}(\vb{m}_{\footnotesize \mbox{1D}})$} \\
		
	\end{tabularx}
	
	\caption{Conceptual visualization of the different forward models. The input is either a multidimensional subsurface model (B. and C.) or an 1D subsurface model (A. and D.). The forward model is either a Low-Fidelity (LF) analytical forward model (with depths and EC as input) (see A.), a High-Fidelity (HF) forward simulation on an accurate mesh (B.) or a Medium- Fidelity (MF) forward simulation on an inaccurate (coarser) mesh (C. and D.). NOTE: the presented meshes are illustrative and are not the ones used for multidimensional modelling. Adapted from \citet{deleersnyder2022novel}.}
	\label{fig:three_models} 
\end{figure}

The accuracy of such HF TDEM simulations is usually limited due to the computational burden, and the response contains a modelling error. The modelling error is a discretization error due to a spatial and/or temporal discretization that is too coarse. We will, however, still utilize those imperfect simulations $\mathcal{F}_{\footnotesize  \mbox{3D, MF}}$ and refer to this model as the medium-fidelity model (MF). We visualize the three types of modelling in Figure \ref{fig:three_models}. This work uses the finite volume method from the open-source package SimPEG \citep{heagy2017framework} for the numerical simulations.\\

\subsection{Joining models}
We assume that the modelling error of the numerical simulation is mainly a discretization error, which depends on the mesh properties (e.g., spatial and temporal discretization) and the subsurface model $\vb{m}$. The difference between the data from a model $\vb{m}$ and a horizontally layered model $\vb{m}_{\footnotesize \mbox{1D}}$ could eliminate a common discretization error, leaving a discrepancy resulting from a model-dependent discretization error and, more importantly, the difference in subsurface model $\vb{m}$, which is what we want to exploit. 
It means that
\begin{linenomath*}
	\begin{equation}
		\mathcal{F}_{\mbox{\footnotesize 3D}}(\vb{m}) - \mathcal{F}_{\footnotesize  \mbox{1D}}(\vb{m}_{\footnotesize \mbox{1D}}) \approx   \mathcal{F}_{\footnotesize  \mbox{3D, MF}}(\vb{m})  - \mathcal{F}_{\footnotesize  \mbox{3D, MF}}(\vb{m}_{\footnotesize \mbox{1D}}),
	\end{equation}
\end{linenomath*}
and after the following rescaling of the right-hand side $\mathcal{F}_{\footnotesize  \mbox{1D}}(\vb{m}_{\footnotesize \mbox{1D}})/ \mathcal{F}_{\footnotesize  \mbox{3D, MF}}( \vb{m}_{\footnotesize \mbox{1D}} )$, we find
\begin{linenomath*}
	\begin{equation}
		\mathcal{F}_{\mbox{\footnotesize 3D}}(\vb{m})  \approx \mathcal{F}_{\footnotesize  \mbox{1D}}(\vb{m}_{\footnotesize \mbox{1D}}) +  \frac{\mathcal{F}_{\footnotesize  \mbox{1D}}(\vb{m}_{\footnotesize \mbox{1D}})}{ \mathcal{F}_{\footnotesize  \mbox{3D, MF}}( \vb{m}_{\footnotesize \mbox{1D}} )} \left( \mathcal{F}_{\footnotesize  \mbox{3D, MF}}(\vb{m})  - \mathcal{F}_{\footnotesize  \mbox{3D, MF}}(\vb{m}_{\footnotesize \mbox{1D}})\right).
	\end{equation}
\end{linenomath*}
We find that the HF data can be predicted $\tilde{\mathcal{F}}_{\mbox{\footnotesize 3D}}(\vb{m})$ by evaluating the LF model and a correction term. Simplifying yields
\begin{linenomath*}
	\begin{equation}
		\label{eq:approach}
		\tilde{\mathcal{F}}_{\mbox{\footnotesize 3D}}(\vb{m}) = 
		\frac{\mathcal{F}_{\footnotesize  \mbox{1D}}(\vb{m}_{\footnotesize \mbox{1D}})}{\mathcal{F}_{\footnotesize  \mbox{3D, MF}}(\vb{m}_{\footnotesize \mbox{1D}}) } \mathcal{F}_{\footnotesize  \mbox{3D, MF}}(\vb{m}),
	\end{equation}
\end{linenomath*}
which avoids evaluating the computationally expensive HF mesh. If the MF mesh would produce accurate results, the ratio becomes 1. Evaluating a 1D model eliminates the contribution of the MF mesh and ensures accurate results.\\

In model order reduction terminology, we combine two different types of hierarchical surrogates from the high-fidelity model, one model based on simplifying physics assumptions (1D assumption) and coarser grids \citep{benner2015survey}.  

\section{Results}
We first show the motivating example behind the proposed method to demonstrate how much the modelling error and computational cost can be reduced. Then, in Sections \ref{sec:effect_model} and \ref{sec:effect_discretization}, we investigate the effect of the subsurface model and the discretization on the accuracy of the prediction, respectively.
\subsection{Example}

\begin{figure}
	\centering
	\includegraphics[width=0.6\textwidth]{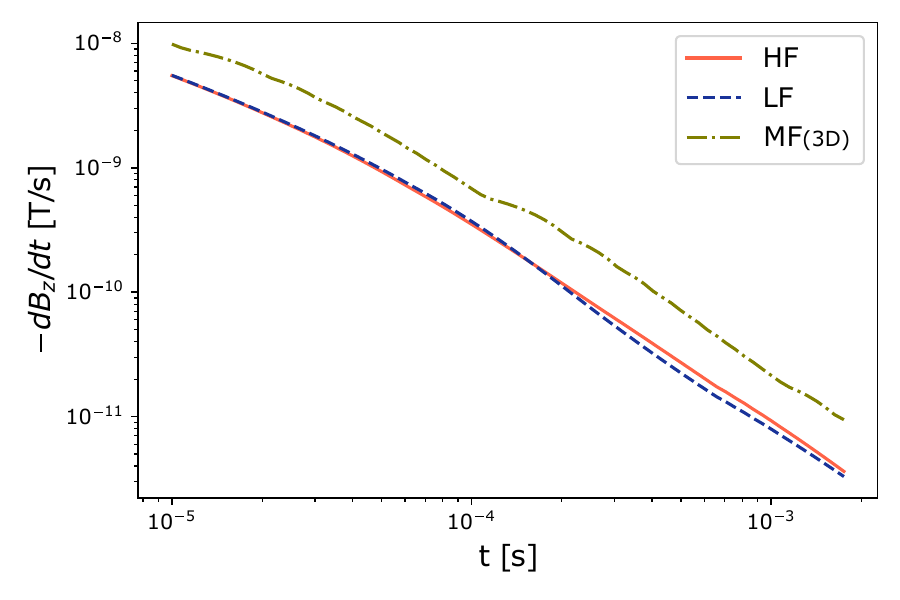}\\
	\includegraphics[width=0.6\textwidth]{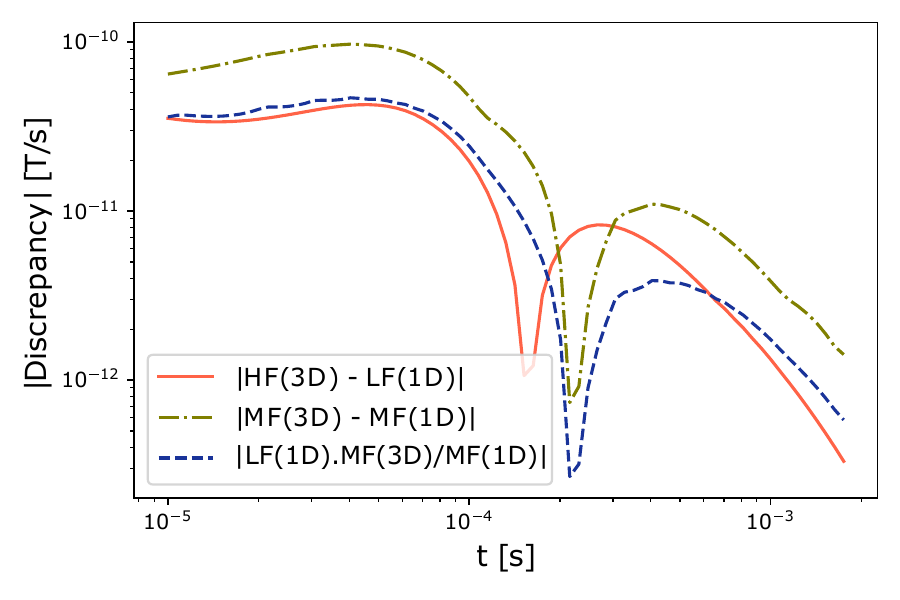}
	\caption{(\textbf{upper}) Data for a two-layered model with a linear interface at depth 50 m below the receiver and an angle of 25$^\circ$ from the three types of forward model. \textbf{(lower)} Discrepancy between a two-layered model with ($\vb{m}$) and without ($\vb{m}_{\footnotesize \mbox{1D}}$) angle.}
	\label{fig:example}
\end{figure}

Consider a two-layered model $\vb{m}$ with a straight interface that can vary at an angle $\theta$ with respect to purely horizontal layers ($\theta = 0$). Evaluating this model $\vb{m}$ in the three forward models $ \mathcal{F}_{\footnotesize  \mbox{1D}}$, $ \mathcal{F}_{\footnotesize  \mbox{3D,MF}}$, $ \mathcal{F}_{\footnotesize  \mbox{3D,HF}}$, yields the magnetic data shown in Figure \ref{fig:example}. The forward models mimic the SkyTEM AEM system, with the source modelled as a vertical magnetic dipole at 40 m altitude. The receiver measures the $z$-component of the magnetic field and is located 2 m higher and is 13.2 m off-centre. The LF model is 1D. Thus, it ignores the angle $\theta$. The MF mesh is a 3D mesh and consists of $36,120$ cells, with the smallest cells in the centre measuring 5x5 meters and gradually increasing in size. The spatial extent is 800 m wide and 300 m deep. There are only 60 time steps. The HF mesh consists of $288,000$ cells; the smallest cells are 1x1 m, and the spatial extent is 4.5 km wide and 1.25 km deep. At least 300 time steps are needed to limit the discretization error. Details about the meshes are provided in supplementary materials. The LF and HF data in Figure \ref{fig:example} are very similar, and the modelling error is mainly due to the difference in the subsurface model \citep{deleersnyder2022novel}. The MF data is inaccurate, so the LF forward model would be more suitable to interpret the data.\\

If we evaluate the same model as in the LF model, namely $\vb{m}_{\footnotesize \mbox{1D}}$, in the MF mesh and look at the difference between the data, namely $|	\mathcal{F}_{\mbox{\footnotesize 3D}}(\vb{m}) - \mathcal{F}_{\footnotesize  \mbox{1D}}(\vb{m}_{\footnotesize \mbox{1D}})|$ and $|\mathcal{F}_{\footnotesize  \mbox{3D, MF}}(\vb{m})  - \mathcal{F}_{\footnotesize  \mbox{3D, MF}}(\vb{m}_{\footnotesize \mbox{1D}}) |$, we observe a similar trend. The minimum has shifted somewhat and there remains a discrepancy in absolute terms. However, using $|\mathcal{F}_{\footnotesize  \mbox{3D, MF}}(\vb{m})  - \mathcal{F}_{\footnotesize  \mbox{3D, MF}}(\vb{m}_{\footnotesize \mbox{1D}}) |$ and normalizing with $N = \mathcal{F}_{\footnotesize  \mbox{1D}}(\vb{m}_{\footnotesize \mbox{1D}})/ \mathcal{F}_{\footnotesize  \mbox{3D, MF}}( \vb{m}_{\footnotesize \mbox{1D}} )$, we find a good approximation of the modelling error using the MF mesh but with a significantly reduced computational cost compared to an HF forward model evaluation.  The HF forward model requires 1090 seconds and 18.5 GB on a HPC, while the MF mesh only requires 20 seconds runtime and 1.25 GB. For conductivities below 0.1 S/m, a HF mesh with more cells will be required to guarantee accurate results, which results in higher costs or more time savings with the MF-approach.

\subsection{Effect of subsurface model}
\label{sec:effect_model}
\begin{figure}
	\centering
	\includegraphics[width=0.6\textwidth]{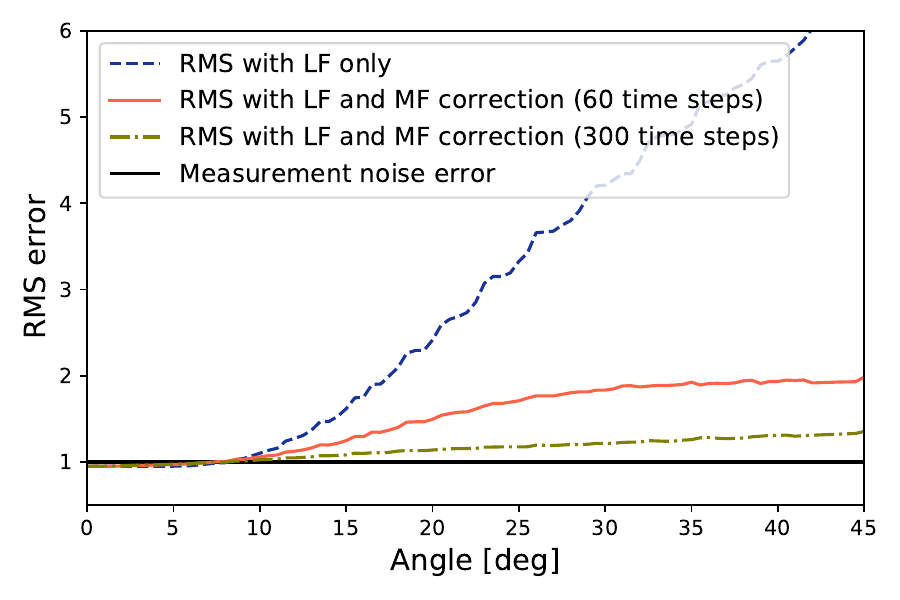}
	\caption{Effect of the angle. The RMS error between Gaussian noisy data and predicted data with three different models: the LF (1D) model, the joined model with a reduced time discretization and the joined model with a finer time discretization. From modelling errors at angles larger than 10°, we can take advantage of mixing the LF model with coarse grids.}
	\label{fig:anglesweep}
\end{figure}

In this section, we explore to what extent the accuracy of the proposed method depends on the subsurface model $\vb{m}$. We compute the root mean squared error
\begin{equation}
	\mbox{RMS} = \sqrt{\frac{1}{n_d} \sum_{i}^{n_d}\left( \frac{\mathcal{F}_{\mbox{\footnotesize 3D, HF}}(\vb{m})_i-  \tilde{\mathcal{F}}_{\mbox{\footnotesize 3D}}(\vb{m})_i }{0.03 \mathcal{F}_{\mbox{\footnotesize 3D, HF}}(\vb{m})_i} \right)^2 },
\end{equation}
for models $\vb{m}$ with increasing angle, thus deviating more from the 1D assumption. 3\% multiplicative Gaussian noise is added to the HF data to mimic measurement noise. If the RMS error approximately equals one, the remaining modelling noise of $\tilde{\mathcal{F}}_{\mbox{\footnotesize 3D}}(\vb{m})$ is difficult to distinguish from the measurement noise.  The results are shown in Figure \ref{fig:anglesweep}.\\

Consider a model $\vb{m}$ with 0.1 and 0.4 S/m for the upper and lower layer respectively and an interface at 50 m depth. For small angles, the modelling error seems irrelevant compared to the measurement noise level (however, it still may affect the interpretation). Starting from a 10$^\circ$  deviation, the 1D LF forward model introduces a significant modelling error - the approach of Eq. \eqref{eq:approach} reduces the modelling error. We considered two MF forward models with different temporal discretization, which affects the accuracy.\\

\begin{figure}
	\centering
	\includegraphics[width=0.6\textwidth]{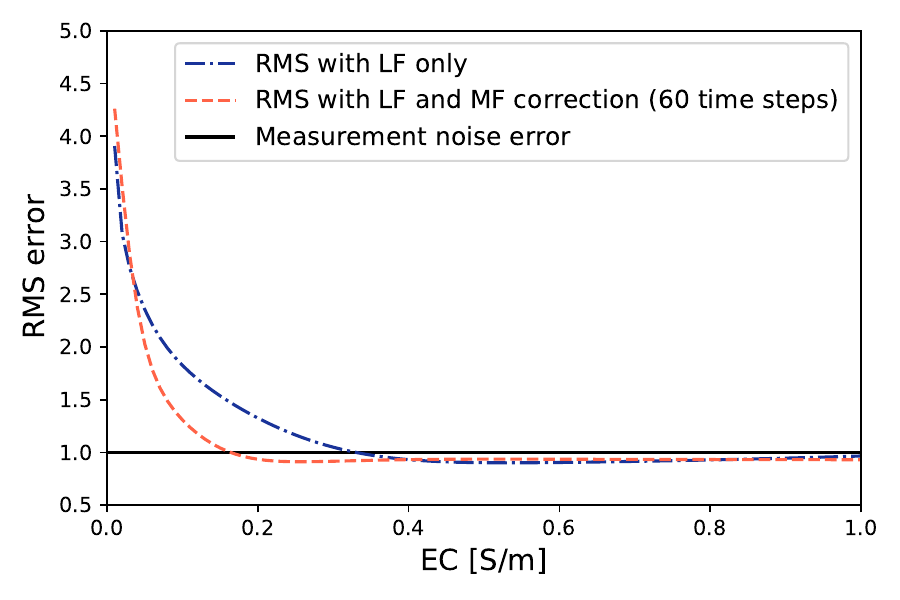}
	\caption{Effect of the EC of the top layer. The RMS error between Gaussian noisy data and predicted data with two different models: the LF (1D) model and the joined model with a reduced time discretization. Varying the EC of the upper layer yields different modelling errors. We can take advantage of joining models for EC between 0.1 and 0.4 S/m. The EC of the lower layer is 0.4 S/m.}
	\label{fig:ECsweep}
\end{figure}

When varying the conductivity of the first layer with an angle of 15°, we observe in Figure \ref{fig:ECsweep} that the modelling error increases significantly for small conductivities. For very small conductivities, we do not longer gain from the approach of Eq. \eqref{eq:approach}. Changing the spatial and temporal discretizations may resolve this. 

\subsection{Effect of temporal discretization}
\label{sec:effect_discretization}
\begin{figure}
	\centering
	\includegraphics[width=0.6\textwidth]{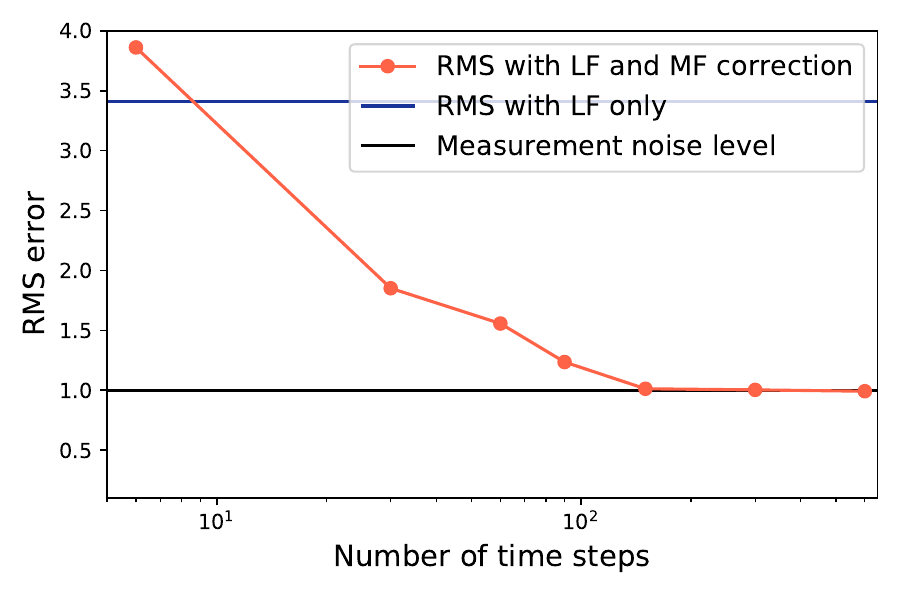}
	\caption{Effect of the temporal discretization. The RMS error between Gaussian noisy data and predicted data with two different models: the LF (1D) model and the joined model with a reduced time discretization. Varying the number of times steps in the coarse discretization yields different modelling errors. We can take advantage of joining models for a minimal number of 30 time steps for this considered subsurface model.}
	\label{fig:temporal_discretization}
\end{figure}

In this section, we explore to what extent the accuracy of the proposed method depends on the discretization of the MF mesh. We limit this work to the temporal discretization. The results are shown in Figure \ref{fig:temporal_discretization}.\\

As expected, the accuracy improves as more timesteps are used. For 30 timesteps, the RMS error is halved compared to when one solely relies on the LF forward model. From 150 timesteps onwards, the modelling error becomes negligible compared to the measurement noise for the considered model $\vb{m}$.


\section{Discussion}
Our proposed approach is an alternative for computationally expensive (time-domain) 3D simulations if an analytical 1D forward model is known, thereby reducing the energy footprint. For time-domain AEM, it allows for more advanced applications than possible with the commonly used 1D forward model. It has the potential to make 3D inversion more available to everyone, as well as to those who do not have access to advanced HPC infrastructures. While future optical computing solutions promise better speed and energy efficiency \citep{xue2024fully}, making large-scale 3D inversions feasible, the availability and cost of the technology remain uncertain. Therefore, it is still worthwhile to study these alternative approaches.\\

The preliminary results of this study are promising. It shows that the absolute value of the MF simulations does not need to be accurate. The disadvantage is that it requires two MF mesh evaluations. The mesh with 60 time steps is 27 times faster and uses much less memory than the HF mesh. It remains an open question of how the performance of the MF mesh can be estimated a priori. We propose conducting a benchmark with an HF mesh in the relevant prior range of your application. At first glance, a relatively simple discretization is sufficient. Low computation costs of the MF mesh ensure that tuning the discretization is a quick process, if required.\\

With the proposed approach, a residual modelling error will remain. This may still cause problems, even for an RMS error below one and when the RMS error criterion assumes Gaussian noise. While the effect of the modelling errors on the inversion is not sufficiently studied today, we assume that minimizing the modelling error will only be beneficial compared to prevailing approaches solely relying on the 1D forward model. 
Note that with this method, we do not need large pre-computed high-fidelity 3D datasets, unlike surrogate models or other AI solutions.\\

\noindent We expect the impact of these results in three domains:
\begin{enumerate}
	\item By looking at the difference of the data on an imperfect MF mesh, we can investigate the effect of modelling errors due to using simplified physics models on the inversion, as demonstrated earlier in \citet{deleersnyder2022novel}.
	\item With Eq.~\eqref{eq:approach}, we can achieve a significant speed-up for 3D deterministic inversion and limit the number of free parameters. The forward modelling does not guarantee exact results. The implications of the residual modelling error are still unclear and need further investigation.
	\item Stochastic interpretations require significantly more evaluations of the forward model, making stochastic 3D interpretations impossible to date. While our approach is not (yet) exact, this is generally not needed for stochastic approaches, which can account for modelling errors \citep{hansen2014accounting}. We hypothesize that such a statistical modelling error relation is less prone to the residual modelling error, which Eq.~\eqref{eq:approach} can quickly generate. 
\end{enumerate}
We have proposed a method with a significant potential, but future research should further validate the approach and reveal whether a more thoughtful spatial and temporal discretization can further reduce the number of cells of the mesh, for example with automatic grid coarsening approaches \citep{werthmuller2019emg3d}. Preliminary results suggest that the spatial discretization near interfaces should not be too coarsened. Furthermore, a better criterion is required to determine whether an MF mesh is sufficient or optimal, whereby accuracy and computation costs are balanced against each other.
\section{Conclusion}
In this study, we have developed a novel approach that integrates 1D simplified physics models with coarse 3D grid simulations to enhance the efficiency of 3D forward modelling. We have exemplified the approach of time-domain 3D AEM data.
The proposed method leverages the accuracy of 1D models while accommodating the flexibility of 3D simulations, significantly reducing modelling errors and computational costs. Our results demonstrate that combining these models can achieve full 3D data with significantly reduced modelling errors but with fewer computational resources. This approach addresses a critical need for faster, more efficient modelling techniques in geophysical applications, particularly as the demand for large-scale subsurface characterization grows in fields such as hydrogeological mapping.\\

The promising results obtained in this work suggest several directions for future research. One potential avenue is to combine the results with multigrid approaches, which automatically use grid coarsening.

\section*{Acknowledgments}

The research leading to these results has received funding from the KU Leuven Postdoctoral Mandate PDMT2/23/065. The authors declare no conflicts of interest.


\begin{thebibliography}{}

\end{thebibliography}


\begin{thebibliography}{}
	
	\bibitem[Ansari et~al., 2017]{ansari2017gauged}
	Ansari, S., Farquharson, C., and MacLachlan, S. (2017).
	\newblock A gauged finite-element potential formulation for accurate inductive
	and galvanic modelling of 3-{D} electromagnetic problems.
	\newblock {\em Geophysical Journal International}, 210(1):105--129.
	
	\bibitem[Asif et~al., 2022]{asif2022integrating}
	Asif, M.~R., Foged, N., Maurya, P.~K., Grombacher, D.~J., Christiansen, A.~V.,
	Auken, E., and Larsen, J.~J. (2022).
	\newblock Integrating neural networks in least-squares inversion of airborne
	time-domain electromagnetic data.
	\newblock {\em Geophysics}, 87(4):E177--E187.
	
	\bibitem[Auken et~al., 2017]{auken2017review}
	Auken, E., Boesen, T., and Christiansen, A.~V. (2017).
	\newblock A review of airborne electromagnetic methods with focus on
	geotechnical and hydrological applications from 2007 to 2017.
	\newblock {\em Advances in Geophysics}, 58:47--93.
	
	\bibitem[Benner et~al., 2015]{benner2015survey}
	Benner, P., Gugercin, S., and Willcox, K. (2015).
	\newblock A survey of projection-based model reduction methods for parametric
	dynamical systems.
	\newblock {\em SIAM review}, 57(4):483--531.
	
	\bibitem[Bording et~al., 2021]{bording2021machine}
	Bording, T.~S., Asif, M.~R., Barfod, A.~S., Larsen, J.~J., Zhang, B.,
	Grombacher, D.~J., Christiansen, A.~V., Engebretsen, K.~W., Pedersen, J.~B.,
	Maurya, P.~K., et~al. (2021).
	\newblock Machine learning based fast forward modelling of ground-based
	time-domain electromagnetic data.
	\newblock {\em Journal of Applied Geophysics}, 187:104290.
	
	\bibitem[B{\"o}rner et~al., 2015]{borner2015three}
	B{\"o}rner, R.-U., Ernst, O.~G., and G{\"u}ttel, S. (2015).
	\newblock Three-dimensional transient electromagnetic modelling using rational
	{K}rylov methods.
	\newblock {\em Geophysical Journal International}, 202(3):2025--2043.
	
	\bibitem[Castillo-Reyes et~al., 2023]{castilloreyes2023meshing}
	Castillo-Reyes, O., Rulff, P., Schankee~Um, E., and Amor-Martin, A. (2023).
	\newblock Meshing strategies for 3d geo-electromagnetic modeling in the
	presence of metallic infrastructure.
	\newblock {\em Computational Geosciences}, 27(6):1023--1039.
	
	\bibitem[Christensen, 2002]{christensen2002generic}
	Christensen, N.~B. (2002).
	\newblock A generic 1-{D} imaging method for transient electromagnetic data.
	\newblock {\em Geophysics}, 67(2):438--447.
	
	\bibitem[Cockett et~al., 2015]{cockett2015simpeg}
	Cockett, R., Kang, S., Heagy, L.~J., Pidlisecky, A., and Oldenburg, D.~W.
	(2015).
	\newblock {SimPEG}: An open source framework for simulation and gradient based
	parameter estimation in geophysical applications.
	\newblock {\em Computers {\&} Geosciences}, 85:142--154.
	
	\bibitem[Deleersnyder et~al., 2022]{deleersnyder2022novel}
	Deleersnyder, W., Dudal, D., and Hermans, T. (2022).
	\newblock Novel {A}irborne {EM} {I}mage {A}ppraisal {T}ool for {I}mperfect
	{F}orward {M}odeling.
	\newblock {\em Remote Sensing}, 14(22):5757.
	
	\bibitem[Deleersnyder et~al., 2024]{deleersnyder2024multidimensional}
	Deleersnyder, W., Dudal, D., and Hermans, T. (2024).
	\newblock A multidimensional ai-trained correction to the 1d approximate model
	for airborne tdem sensing.
	\newblock {\em Computers \& Geosciences}, 188:105602.
	
	\bibitem[Deleersnyder et~al., 2023]{deleersnyder2023flexible}
	Deleersnyder, W., Maveau, B., Hermans, T., and Dudal, D. (2023).
	\newblock Flexible quasi-2{D} inversion of time-domain {AEM} data, using a
	wavelet-based complexity measure.
	\newblock {\em Geophysical Journal International}, 233(3):1847--1862.
	
	\bibitem[Engebretsen et~al., 2022]{engebretsen2022accelerated}
	Engebretsen, K.~W., Zhang, B., Fiandaca, G., Madsen, L.~M., Auken, E., and
	Christiansen, A.~V. (2022).
	\newblock Accelerated 2.5-{D} inversion of airborne transient electromagnetic
	data using reduced 3-{D} meshing.
	\newblock {\em Geophysical Journal International}, 230(1):643--653.
	
	\bibitem[Guillemoteau et~al., 2012]{guillemoteau2012fast}
	Guillemoteau, J., Sailhac, P., and Behaegel, M. (2012).
	\newblock Fast approximate 2{D} inversion of airborne {TEM} data: Born
	approximation and empirical approach.
	\newblock {\em Geophysics}, 77(4):WB89--WB97.
	
	\bibitem[Haber et~al., 2007]{haber2007inversion}
	Haber, E., Oldenburg, D.~W., and Shekhtman, R. (2007).
	\newblock Inversion of time domain three-dimensional electromagnetic data.
	\newblock {\em Geophysical Journal International}, 171(2):550--564.
	
	\bibitem[Hansen et~al., 2014]{hansen2014accounting}
	Hansen, T.~M., Cordua, K.~S., Jacobsen, B.~H., and Mosegaard, K. (2014).
	\newblock Accounting for imperfect forward modeling in geophysical inverse
	problems - exemplified for crosshole tomography.
	\newblock {\em {GEOPHYSICS}}, 79(3):H1--H21.
	
	\bibitem[Heagy et~al., 2017]{heagy2017framework}
	Heagy, L.~J., Cockett, R., Kang, S., Rosenkjaer, G.~K., and Oldenburg, D.~W.
	(2017).
	\newblock A framework for simulation and inversion in electromagnetics.
	\newblock {\em Computers {\&} Geosciences}, 107:1--19.
	
	\bibitem[Hunziker et~al., 2015]{hunziker2015electromagnetic}
	Hunziker, J., Thorbecke, J., and Slob, E. (2015).
	\newblock The electromagnetic response in a layered vertical transverse
	isotropic medium: A new look at an old problem.
	\newblock {\em Geophysics}, 80(1):F1--F18.
	
	\bibitem[Ikard et~al., 2023]{ikard2023model}
	Ikard, S.~J., Minsley, B.~J., Rigby, J.~R., and Kress, W.~H. (2023).
	\newblock A model of transmissivity and hydraulic conductivity from electrical
	resistivity distribution derived from airborne electromagnetic surveys of the
	mississippi river valley alluvial aquifer, midwest usa.
	\newblock {\em Hydrogeology Journal}, 31(2):313--334.
	
	\bibitem[Ley-Cooper, 2022]{leycooper2022acquisition}
	Ley-Cooper, Y. (2022).
	\newblock Acquisition of airborne electromagnetic data at a continental scale.
	\newblock In {\em NSG2022 3rd Conference on Airborne, Drone and Robotic
		Geophysics}, volume 2022, pages 1--5. European Association of Geoscientists
	\& Engineers.
	
	\bibitem[Liu et~al., 2017]{liu2017wavelet}
	Liu, Y., Farquharson, C.~G., Yin, C., and Baranwal, V.~C. (2017).
	\newblock Wavelet-based 3-{D} inversion for frequency-domain airborne {EM}
	data.
	\newblock {\em Geophysical Journal International}, 213(1):1--15.
	
	\bibitem[Mikucki et~al., 2015]{mikucki2015deep}
	Mikucki, J.~A., Auken, E., Tulaczyk, S., Virginia, R., Schamper, C.,
	S{\o}rensen, K., Doran, P., Dugan, H., and Foley, N. (2015).
	\newblock Deep groundwater and potential subsurface habitats beneath an
	{A}ntarctic dry valley.
	\newblock {\em Nature communications}, 6(1):1--9.
	
	\bibitem[Oldenburg and Ellis, 1991]{oldenburg1991inversion}
	Oldenburg, D. and Ellis, R. (1991).
	\newblock Inversion of geophysical data using an approximate inverse mapping.
	\newblock {\em Geophysical Journal International}, 105(2):325--353.
	
	\bibitem[Pfaffhuber et~al., 2017]{pfaffhuber2017delineating}
	Pfaffhuber, A.~A., Lysdahl, A.~O., S{\o}rmo, E., Skurdal, G.~H., Thomassen, T.,
	Ansch{\"u}tz, H., and Scheibz, J. (2017).
	\newblock Delineating hazardous material without touching - {AEM} mapping of
	{N}orwegian alum shale.
	\newblock {\em First Break}, 35(8).
	
	\bibitem[Podgorski et~al., 2013]{podgorski2013processing}
	Podgorski, J.~E., Auken, E., Schamper, C., Vest~Christiansen, A., Kalscheuer,
	T., and Green, A.~G. (2013).
	\newblock Processing and inversion of commercial helicopter time-domain
	electromagnetic data for environmental assessments and geologic and
	hydrologic mapping.
	\newblock {\em Geophysics}, 78(4):E149--E159.
	
	\bibitem[Puzyrev and Swidinsky, 2021]{puzyrev2021inversion}
	Puzyrev, V. and Swidinsky, A. (2021).
	\newblock Inversion of 1{D} frequency-and time-domain electromagnetic data with
	convolutional neural networks.
	\newblock {\em Computers \& Geosciences}, 149:104681.
	
	\bibitem[Rulff et~al., 2021]{rulff2021efficient}
	Rulff, P., Buntin, L.~M., and Kalscheuer, T. (2021).
	\newblock Efficient goal-oriented mesh refinement in 3-{D} finite-element
	modelling adapted for controlled source electromagnetic surveys.
	\newblock {\em Geophysical Journal International}, 227(3):1624--1645.
	
	\bibitem[Schamper et~al., 2014]{schamper2014assessment}
	Schamper, C., J{\o}rgensen, F., Auken, E., and Effers{\o}, F. (2014).
	\newblock Assessment of near-surface mapping capabilities by airborne transient
	electromagnetic data—an extensive comparison to conventional borehole data.
	\newblock {\em Geophysics}, 79(4):B187--B199.
	
	\bibitem[Siemon et~al., 2019]{siemon2019automatic}
	Siemon, B., van Baaren, E., Dabekaussen, W., Delsman, J., Dubelaar, W.,
	Karaoulis, M., and Steuer, A. (2019).
	\newblock Automatic identification of fresh -- saline groundwater interfaces
	from airborne electromagnetic data in {Z}eeland, the {N}etherlands.
	\newblock {\em Near Surface Geophysics}, 17(1):3--25.
	
	\bibitem[Werthm{\"u}ller et~al., 2019]{werthmuller2019emg3d}
	Werthm{\"u}ller, D., Mulder, W.~A., and Slob, E.~C. (2019).
	\newblock emg3d: A multigrid solver for 3{D} electromagnetic diffusion.
	\newblock {\em Journal of Open Source Software}, 4(39):1463.
	
	\bibitem[Werthmüller, 2017]{werthmueller2017open}
	Werthmüller, D. (2017).
	\newblock An open-source full 3{D} electromagnetic modeler for 1{D} {VTI} media
	in {P}ython: empymod.
	\newblock {\em {GEOPHYSICS}}, 82(6):WB9--WB19.
	
	\bibitem[Wolfgram et~al., 2003]{wolfgram2003approximate}
	Wolfgram, P., Sattel, D., and Christensen, N.~B. (2003).
	\newblock Approximate 2{D} inversion of {AEM} data.
	\newblock {\em Exploration Geophysics}, 34(2):29--33.
	
	\bibitem[Wu et~al., 2023]{wu2023deep}
	Wu, S., Huang, Q., and Zhao, L. (2023).
	\newblock A deep learning-based network for the simulation of airborne
	electromagnetic responses.
	\newblock {\em Geophysical Journal International}, 233(1):253--263.
	
	\bibitem[Xue et~al., 2024]{xue2024fully}
	Xue, Z., Zhou, T., Xu, Z., Yu, S., Dai, Q., and Fang, L. (2024).
	\newblock Fully forward mode training for optical neural networks.
	\newblock {\em Nature}, 632(8024):280--286.
	
	\bibitem[Yang and Oldenburg, 2017]{yang20173d}
	Yang, D. and Oldenburg, D.~W. (2017).
	\newblock 3d inversion of total magnetic intensity data for time-domain em at
	the lalor massive sulphide deposit.
	\newblock {\em Exploration Geophysics}, 48(2):110--123.
	
\end{thebibliography}
\end{document}